\documentstyle[12pt]{article}

\begin{document}
\thispagestyle{empty}

\begin{center}
{\large \bf
A partition functional and thermodynamic properties  \\
    of  the infinite-dimensional Hubbard model }\\
\bigskip
{\bf Y. M. Li}    and   {\bf N. d'Ambrumenil} \\
\smallskip
{\it Department of Physics, University of Warwick,
                            Coventry CV4 7AL, United  Kingdom}\\
\end{center}

\bigskip

\begin{abstract}
  An approximate  partition functional is derived
for the infinite-dimensional Hubbard model.
This functional naturally includes the exact solution of the
Falicov-Kimball model as a special case, and is
exact  in the uncorrelated and  atomic limits.
It explicitly keeps spin-symmetry.
For the case of  the Lorentzian density of states,
we find that the Luttinger theorem is satisfied at zero temperature.
The susceptibility
crosses over smoothly from that expected for an  uncorrelated state with
antiferromagnetic fluctuations
at high temperature to a correlated state at
low temperature via a Kondo-type anomaly at a characteristic
temperature $T^\star$.  We attribute this anomaly   to
the appearance of the  Hubbard pseudo-gap.  The  specific heat also  shows
a peak  near  $T^\star$.
The resistivity goes to zero at zero temperature,  in contrast
to other approximations,  rises sharply around  $T^\star$
and has a rough linear temperature dependence above $T^\star$.
\end{abstract}

\bigskip
PACS numbers: 71.28.+d,  75.20.Hr

\newpage

\section{Introduction}

The single-band Hubbard model (HM) and the
periodic Anderson model have  attracted a great deal of interest since
the heavy fermion and high temperature superconductors were discovered.
Although there has been intensive theoretical work on both models,
the knowledge we have for them  is still limited,
with an exact solution known only for  the one-dimensional HM \cite{LW}.
An important advance was made by Metzner and Vollhardt \cite{MV}, who
observed  that   any diagrammatic treatment of
fermion lattice models  simplifies in infinite dimensions.
The proper  self-energy becomes site-diagonal and  momentum
conservation is thus irrelevant \cite{Muller}.  Despite this simplification,
the dynamic properties remain  non-trivial.  This discovery has stimulated
much attention recently  \cite{Rev}.
Nearly all the phenomena associated with strong electron correlations have
been found in the infinite dimensional models:
the Mott transition \cite{MT},  antiferromagnetism
\cite{Jarrell}, Fermi liquid  or
non-Fermi liquid  behavior \cite{Si93}, and even the possibility of
superconductivity \cite{SC}. The infinite dimensional models appear to be
a good starting point for studying
electron correlations.

Brandt and Mielsch  \cite{BM} first showed that
the infinite dimensional (${D^{\infty}}$)
HM  can be mapped onto an atomic
problem in the presence of two time-dependent external fields.
They were able to solve exactly
the ${D^{\infty}}$ Falicov-Kimball model, which
corresponds to one time-dependent external field.
A complete exact solution for
the  case of two external fields  has not been possible.
An equivalent mapping to the Anderson impurity  model with a self-consistent
condition  has also been  found \cite{GK, Jarrell}.
This has been used to calculate properties of the
${D^{\infty}}$ HM \cite{MT, Jarrell} and
periodic Anderson model  \cite{Jarrell2} by the
quantum Monte Carlo (QMC) method.
Although the QMC method is essentially
exact,  it can not  treat  the whole parameter region,
in particular the cases of large $U$ and low temperatures.
Analytical continuation using the maximum entropy method to
calculate some dynamic quantities is also problematic.
A controlled approximate solution is clearly desirable for the whole
parameter region.

Many approximate schemes have
been developed  for the strongly correlated electron systems, such as
the alloy analogy approximation (plus coherent potential
approximation),  equation-of-motion decoupling \cite{Czy},
slave boson \cite{Boson}  and slave fermion \cite{Fermion} approaches,
{\it etc.}.
Unfortunately, they are usually not controlled approximations and
do not become exact in
physically-relevant limits.

In the present paper  we report on a better approximation scheme
for the infinite dimensional models.
The alloy analogy approximation \cite{Czy} (as well as  the Hubbard-III
approximation \cite{Hubbard-III} and higher-order equation-of-motion
decoupling \cite{Czy}) do rather well for high temperatures,
and are exact in the
free electron and  atomic limits. But they  violate the Luttinger theorem
at $T=0$, and do not include magnetic order probably
as a result of breaking  spin
symmetry.  The slave-boson and slave-fermion mean-field approximations
usually give us contradictory results, and do not become exact
even in the limit of free electrons.  Our approximation scheme  retains
the merits of  the previous approximations especially at high temperature
and respects the spin symmetry explicitly.
It satisfies the Luttinger theorem at zero temperature thereby remedying
the most
serious defect of   approximations such as the alloy analogy
approximation. Furthermore,   the
approximate solution is obtained in
an expansion scheme so that it can be improved systematically by
including higher-order effects.

Jani\u{s} and Vollhardt \cite{Janis} have already suggested
an approximate solution to the ${D^{\infty}}$ Hubbard model.  Their
ansatz for the thermodynamic potential is the
linear combination of the potentials for two Falicov-Kimaball models
obtained   by
assigning one potential to one spin and retaining the other spin
as  the impurity.  Although this ansatz is
formally written to be spin-symmetric,  the ingredients  they used,
{\it i.e.} the solution of the Falicov-Kimball model,  break the
symmetry intrinsicallly.  It does not
satisfy the Luttinger theorem at zero temperature.

We use the Brandt-Mielsch  mapping  \cite{BM} to find
an approximate
solution to the partition functional of the
${D^{\infty}}$ HM (including the periodic Anderson model) \cite{Li}.
{}From the partition functional one can obtain the
self-energy and all the thermodynamic properties of the
system as well as some correlation functions.
The functional we obtain naturally contains the exact solution of
the ${D^{\infty}}$ Falicov-Kimball model as a special case, and is exact
in the uncorrelated  and atomic  limits.
The spin-symmetry is retained explicitly.

To demonstrate the approximation scheme,
we calculate the thermodynamic properties of the ${D^{\infty}}$ HM
for a Lorentzian density of states (DOS).
For this simple DOS, one can
obtain some analytic expressions for physical quantities.
The main physical  properties of the ${D^\infty}$ HM  (except the
Mott transition)
are expected to  remain for this  DOS \cite{Si}.
We find  that at zero temperature the system
is a Fermi liquid (except at  half-filling), which is
consistent with the exact result of Georges and Kotliar  \cite{GK}.
The Luttinger theorem is satisfied.
As the temperature is increased, the system crosses over continuously from a
correlated  state  to an uncorrelated state with a
Curie-Weiss susceptibility via a Kondo-like anomaly.   This is
in contrast to  other approximations in which either  the
local moment behavior observed at high temperatures or
the strongly correlated behavior observed at low temperatures is not properly
reproduced \cite{Temp} \cite{Czy}.  The Kondo-like anomaly here
actually results from
the Hubbard pesudo-gap.  The specific heat shows a peak at the
corresponding (Kondo) characteristic temperature $T^\star$.
The resistivity goes to zero at $T=0$. This is completely different from
the result of the alloy analogy approximation and other schemes,
which break the
Luttinger theorem at zero temperature so that  the
resistivity always goes up as temperature decreases.

The paper is organized as follows.  We briefly outline the
main features of the  ${D^\infty}$ HM and the
Brandt-Mielsch mapping in Section II.  We derive the approximate
partition functional by an  expansion in Section III.
Some physical quantities, such as
the susceptibility, the specific heat and the resistivity
are then calculated  using
the functional in Section IV. Finally, we  discuss our results  in
Section V.

\section{The infinite-dimensional Hubbard model
\newline and equivalent atomic problem}

\subsection{The ${D^\infty}$ Hubbard model}

The  Hubbard Hamiltonian for large dimensions, $D$,
is \cite{MV}
\begin{equation}
  H_H = -\frac{1}{2\sqrt{D}} \sum_{ij \sigma}
          t_{ij} a_{i,\sigma}^\dagger a_{j,\sigma}
     + U \ \sum_{i}  n_{i,\uparrow} n_{i,\downarrow}
      + \sum_{i \sigma} (E_\sigma - \mu) a_{i,\sigma}^\dagger a_{i,\sigma},
\label{e1}
\end{equation}
where $a_{i, \sigma}$, $a_{i, \sigma}^\dagger$ and
$n_{i, \sigma}=$ $a_{i, \sigma}^\dagger a_{i, \sigma}$ are
usual fermion operators;  ${\mu}$ is the chemical potential,  ${E_\sigma}$
takes  account of  the spin-dependence of the
atomic levels. In the presence of the
external field $h$,  ${E_\sigma=E_0-\sigma  \mu_B h}$  ($\mu_B$ the Bohr
magneton).  The hopping term is scaled  so as to make the  kinetic
energy and the interaction term competitive.

Because of the scaling by $1/\sqrt{D}$ for the kinetic term in (\ref{e1}),
the self-energy becomes site-diagonal \cite{MV, Muller},
{\it i.e.} its Fourier transform $\Sigma(i\omega_n, {\bf k})
\equiv \Sigma(i\omega_n)$ is a ${\bf k}$-independent quantity.
For a homogeneous system \cite{Note}, the Dyson equation
for the full momentum-dependent Green's function
of (\ref{e1}) is   given by
\begin{equation}
G_\sigma(i\omega_n,k)= [i\omega_n-(E_\sigma- \mu)-\Sigma_\sigma (i\omega_n)-
\varepsilon(k)]^{-1}.
\label{add1}
\end{equation}
Integrating $G_\sigma(i\omega_n,k)$ over $k$ gives the following local Green's
function
\begin{eqnarray}
G^{\rm loc}_\sigma(i\omega_n)& \equiv &\langle G_n^\sigma (k) \rangle_k
\nonumber \\
&=&\int_{-\infty}^{\infty} \frac{\rho_0(\epsilon) d\epsilon}
{i\omega_n-(E_\sigma- \mu)-\Sigma_\sigma (i\omega_n)
 - \epsilon} .
\label{e8}
\end{eqnarray}
Here ${\rho_0(\epsilon)}$ is the  bare density of states (DOS),  which
is  of  Gaussian type   \cite{MV} \cite{Muller}  for  the nearest
neighbor hopping  in (1):
$$\rho_0(\epsilon)=\sum_k \delta (\epsilon-\epsilon_k)
=\frac{1}{\sqrt{\pi}} {\rm exp} [-\epsilon^2]. $$
The self-energy  can be expressed as a functional of the
local Green's functional $G^{\rm loc}_\sigma(i\omega_n)$,
as shown using the perturbation expansion  in \cite{Muller2}:
\begin{equation}
\Sigma(i\omega_n) =\Sigma[G^{\rm loc}_\sigma(i\omega_n)].
\label{Sigma}
\end{equation}
Once the self-energy functional ( \ref{Sigma}) is known, one can
determine the  local Green's function from
(\ref{e8}), and, consequently, the self-energy.
Eq.\ (\ref{e8}) therefore  provides a self-consistent equation
for determining the Green's function of the system.

\subsection{The Brandt-Mielsch mapping}

The central problem is to find the self-energy, $\Sigma$,
as a functional of the
local Green's function $G^{\rm loc}_\sigma(i\omega_n)$.
Brandt and Mielsch \cite{BM} mapped the  ${D^{\infty}}$ HM onto an
atomic problem with  two external fields to find this
functional.

For an atomic (single site) problem, ${H_H}$ in (1)  reduces to the
atomic Hamiltonian ${H=H_0+V}$,  where
\begin{eqnarray}
  H_0 & =&  \sum_{ \sigma} (E_\sigma - \mu)
          a_{\sigma}^\dagger a_{\sigma}, \nonumber \\
  V  & = &  U \ n_{\uparrow} n_{\downarrow} .
\label{seH}
\end{eqnarray}
Of course, the Hamiltonian $H$ can  easily be  solved.  However, when the
two external fields are introduced, the problem is never trivial.

A generating partition functional is defined as
\begin{equation}
Z={\rm Tr}\{ e^{-\beta H} S \},
\label{e3}
\end{equation}
where
\begin{equation}
S=T_\tau {\rm exp}\{-\int_0^\beta d \tau \int_0^\beta d \tau'
   \sum_\sigma \lambda^\sigma (\tau,\tau') {a_\sigma^H}^\dagger (\tau)
    a_\sigma^H (\tau') \}
\label{e4}
\end{equation}
with ${{a_\sigma^H}^\dagger (\tau)=e^{H\tau} a_\sigma^\dagger e^{-H\tau} }$
and ${\lambda^\sigma (\tau,\tau')}$ is an external field.
The Green's function is defined by functional differentiation:
$\bar{G}_\sigma (\tau, \tau')=-\delta
{\rm ln} Z /\delta \lambda^\sigma (\tau', \tau)$.
Using Matsubara's formalism, we write
\begin{equation}
\lambda^\sigma (\tau,\tau')=\frac{1}{\beta}
\sum_n {\rm exp}[-i \omega_n (\tau-\tau')]
               \lambda_n^\sigma,
\label{e5}
\end{equation}
where ${\omega_n=(2n+1)\pi/\beta}$. The Fourier transform of
the Green's function,
$\bar{G}_\sigma(i \omega_n)$, is then given by
\begin{equation}
\bar{G}_\sigma(i \omega_n)  =  -\partial {\rm ln}
Z/\partial \lambda_n^\sigma
\label{se5}
\end{equation}
The  self-energy  is  defined  via
\begin{equation}
\Sigma_\sigma (i\omega_n) = {G^0_\sigma (i \omega_n)}^{-1} -
               {\bar{G}_\sigma(i \omega_n)}^{-1},
\label{e6}
\end{equation}
where
\begin{equation}
 G^0_\sigma (i \omega_n)  = [i \omega_n -(E_\sigma-\mu)-
 \lambda_n^\sigma]^{-1} .
\label{se7}
\end{equation}

One may ask what the relation is between the local Green's function
$G^{\rm loc}_\sigma(i \omega_n)$ in (\ref{e8}) and the Green's
function $\bar{G}_\sigma(i \omega_n)$ defined in (\ref{se5}).
When one expands the self-energy of (\ref{e6})  in a series of
skeleton diagrams in terms of
$\bar{G}_\sigma(i \omega_n)$, {\it i.e.}, expands in the
formalism of  Baym and Kadanoff \cite{BK},
one finds that there is a one-to-one correspondance
between these skeleton diagrams and those of the self-energy of the
original Hubbard model in (1) in terms of
$G^{\rm loc}_\sigma(i\omega_n)$.
The self-energy (\ref{e6}), when written as a functional of
${\bar{G}_\sigma(i\omega_n)}$, therefore has the same form
as the self-energy functional of the Hubbard model (1).
Since this functional is what is inserted in (\ref{e8}), it
is clear that ${\bar{G}_\sigma(i\omega_n)}$
and ${G^{\rm loc}_\sigma(i\omega_n)}$ are the same
once Eq.\  (\ref{e8}) has been solved self-consistently.

In a real calculation of the local Green's function from the self-consistent
equation (\ref{e8}),  one does not need  to  find an explicit form of the
self-energy as a function of ${\bar{G}_\sigma(i\omega_n)}$.
It is in general easier to determine from (\ref{e8}) the
${\lambda_n^\sigma}$  (or  equivalently
the Green's function ${G^0_\sigma (i \omega_n)}$ of (\ref{se7}) ),
and then to use the  ${\lambda_n^\sigma}$
to fix the self-energy in (\ref{e6}) and ${\bar{G}_\sigma(i\omega_n)}$
in (\ref{se5}), which is equal to
${G^{\rm loc}_\sigma(i\omega_n)}$.  Below we will not differentiate
between ${\bar{G}_\sigma(i\omega_n)}$
and ${G^{\rm loc}_\sigma(i\omega_n)}$ on the
understanding that the external fields are
determined self-consistently.

A complete solution for the partition functional $Z$ in (\ref{e3}) has
not been possible.  The exact solution is known  for the
Falicov-Kimball model \cite{BM}, for which only one of external fields
is needed \cite{FK2} as only  one spin species  is allowed to move
in this model.
In the next section, we  derive an approximate solution  for the
partition functional $Z$ of the ${D^\infty}$ HM.

\section{Approximate partition functional}

\subsection{Expansion}

Since ${H_0}$ and ${V}$   commute
for the atomic Hamiltonian  $H$ in (\ref{seH}),  one can  write
\begin{eqnarray}
e^{-\beta H} &=& e^{-\beta H_0} e^{-\beta V}
   =e^{-\beta H_0} \sum_{m=0}^\infty \frac{(-\beta)^m}{m!} V^m
 \nonumber \\
  & =&   e^{-\beta H_0}  [1+(e^{-\beta U}-1) n_\uparrow n_\downarrow] .
\label{ex}
\end{eqnarray}
The partition functional  of  (\ref{e3}) can therefore  be expressed
\begin{equation}
Z=e^{-\beta \Omega_0} [\langle S \rangle_0 +
    (e^{-\beta U} -1) \langle n_\uparrow n_\downarrow S \rangle_0],
\label{e10}
\end{equation}
where ${\Omega_0}$ is the thermodynamic potential for ${H_0}$,  and
${\langle \theta \rangle_0 ={\rm Tr}(e^{\beta(\Omega_0-H_0) }
\theta) }$.  The important point to note  from  (\ref{e10}) is that
we need only to calculate expectation values with respect to
${H_0}$.  However, the interaction still  appears
in the  Heisenberg operators
${{a_\sigma^H}^\dagger (\tau)}$ and
${{a_\sigma^H} (\tau)}$ which define $S$  in (\ref{e4}).

We  represent the operators ${{a_\sigma^H}^\dagger (\tau)}$ and
${{a_\sigma^H} (\tau)}$ in terms of the Heisenberg operators for
$H_0$ in a similar way as  in (\ref{ex}).  We write
\begin{eqnarray}
  a_\sigma^{H \dagger} (\tau) & = &
  a_\sigma^\dagger(\tau)[1+(e^{U\tau}-1)n_{-\sigma}(\tau)] \nonumber,\\
  a_\sigma^H (\tau) & = &
  a_\sigma(\tau)[1+(e^{-U\tau}-1)n_{-\sigma}(\tau)],
\label{operator}
\end{eqnarray}
here  $a_\sigma^\dagger(\tau) =e^{H_0 \tau} a_\sigma^\dagger e^{-H_0 \tau}$.

 We first look at the expectation value $\langle S \rangle_0$
in (\ref{e10}), which  is   given by
\begin{eqnarray}
\langle S \rangle_0&=&\langle T_{\tau \theta} \{ {\rm exp}
[ -\int_0^\beta d \tau \int_0^\beta d \tau'
   V_\uparrow(\tau,\tau')  ]  \nonumber \\
       && ~~~~~ {\rm exp} [-\int_0^\beta d \theta \int_0^\beta d \theta'
   V_\downarrow(\theta,\theta') ]  \} \rangle_0,
\label{e12}
\end{eqnarray}
where
\begin{eqnarray}
V_\sigma (\tau, \tau') & =& A_\sigma (\tau, \tau') B_{\sigma} (\tau, \tau'),
\nonumber\\
A_\sigma (\tau, \tau') &=& \lambda^\sigma (\tau,\tau') a_\sigma^\dagger(\tau)
                 a_\sigma (\tau'),
\label{VAB1}  \\
B_\sigma (\tau, \tau') & = &  [1+(e^{U\tau}-1) n_{-\sigma}(\tau)]
   [1+(e^{-U\tau'}-1) n_{-\sigma}(\tau')].
\nonumber
\end{eqnarray}
Expanding the  exponentials in (\ref{e12})  gives
\begin{equation}
\langle S \rangle_0=\sum_{mn} \frac{1}{m!n!} \int d \tau
       \int d \theta X_{mn}(\tau, \theta),
\label{SX}
\end{equation}
where
\begin{equation}
 X_{mn}(\tau, \theta)=\langle T_{\tau \theta}\{\bar{A}_\uparrow^m (\tau)
\bar{B}_\uparrow^m (\tau)
             \bar{A}_\downarrow^n (\theta)
\bar{B}_\downarrow^n (\theta) \}  \rangle_0,
\label{add2}
\end{equation}
with $\bar{A}_\sigma^m (\tau)  =
A_\sigma (\tau_1, \tau_1') ... A_\sigma (\tau_m, \tau_m')$,
and
$\bar{B}_\sigma^m (\tau) =
B_\sigma (\tau_1, \tau_1') ... B_\sigma (\tau_m, \tau_m')$, and
the integration in (\ref{SX}) is for all $\tau$'s and $\theta$'s.
$\bar{A}_\sigma^m (\tau)$ and
$\bar{B}_{-\sigma}^n (\tau)$
contain all the operators involving spin-$\sigma$ electrons
in (\ref{add2}), and so
using  Wick's theorem
we can write
\begin{equation}
X_{mn}(\tau, \theta)=\langle T_{\tau \theta}\{\bar{A}_\uparrow^m (\tau)
                \bar{B}_\downarrow^n (\theta) \} \rangle_0
       \langle T_{\tau \theta}\{\bar{A}_\downarrow^n (\theta)
                \bar{B}_\uparrow^m (\tau) \}      \rangle_0.
\label{X}
\end{equation}
The quantity ${X_{mn}}$ may be calculated for any finite order $m$ and
$n$ using  Wick's theorem. However, an analytic calculation up to
infinite order including all  Feynmann diagrams as required in
(\ref{SX}) is not possible.
We therefore approximate to find the ${X_{mn}}$.

\subsection{Approximation}

When evaluating the  ${X_{mn}}$, we will consider only contractions
within  ${\bar{A}_\sigma(\tau)}$
and ${\bar{B}_{-\sigma}(\theta) }$ and ignore those between
${\bar{A}_\sigma(\tau)}$
and  ${\bar{B}_{-\sigma}(\theta) }$.
Note that the effects of the hopping terms in
the original Hamiltonian are incoporated via the external fields,
$\lambda^\sigma (\tau,\tau')$, included in ${\bar{A}_\sigma(\tau)}$
[see Eq.\ (\ref{VAB1})].
Roughly speaking, the
${\bar{A}_\sigma(\tau)}$ account for the ``dynamics'' of the spin $\sigma$
electrons,  while  the ${\bar{B}_{-\sigma}(\theta) }$
monitor the  value of $n_\sigma$ at the times,  $\theta$, at which
$-\sigma$ occupation changes.  Restricting  contractions to within
${\bar{B}_{-\sigma}(\theta) }$  sets up  a mean `$-\sigma$' field
for the motion of the $\sigma$ electrons and {\it vice versa}.
As we will see, $\langle  T_\theta
\bar{B}_{-\sigma}(\theta) \rangle$ gives rise to
two terms which describe propagation in the upper and lower
Hubbard bands.

Our approximation is in  the spirit of the Gutzwiller
approximation for Gutzwiller wave function \cite{Gutzwiller}.
Gutzwiller's original idea was to treat the
down-spin electrons as static when considering  the motion of the
up-spin eletrons, and {\it vice versa}.  This
approximation does not break the spin-symmetry and keeps the
translational invariance, so that  Fermi-liquid behavior can
exist at zero temperature.  The ground state energy per site
for the Hubbard model like Eq.\ (1) can be written as
\begin{equation}
E=\frac{1}{N} \sum_{k, \sigma} q_\sigma \epsilon_k n_{k \sigma} + U d,
\label{EGutzwiller}
\end{equation}
where $n_{k \sigma}$ is the fermion distribution function and
$d$ is the double occupancy per site.
The effects of one spin species on the other are incoporated into
the renormalization factor, $q_\sigma$.
Eq.\  (\ref{EGutzwiller}) has been rederived  statistically
 \cite{Ogata} and the  Gutzwiller approximation has been
shown to be exact in the limit of infinite dimensions \cite{MV}.

Only considering the contractions within ${\bar{A}_\sigma (\tau)}$
and ${\bar{B}_{-\sigma}(\theta) }$ is equivalent to decoupling
${\bar{A}_\sigma(\tau)}$
and ${\bar{B}_{-\sigma}(\theta) }$.  Writing
$\langle T_{\tau}   \bar{A}_\sigma^m (\tau)     \rangle_0 \,
\langle T_{\theta} \bar{B}_{-\sigma}^n (\theta) \rangle_0 $
gives
\begin{equation}
\langle T_{\theta} \bar{B}_{-\sigma}^n (\theta) \rangle_0
=\langle [1-n_{\sigma}] + {\rm exp} \{ U\sum_{i=1}^n
(\theta_i-\theta'_i) \} n_{\sigma}  \rangle_0.
\label{HF}
\end{equation}
Accounting  for the first term of (\ref{HF}), when multiplied by
$\langle T_{\tau} \bar{A}_\uparrow^m  (\tau) \rangle_0$,
is straightforward. Summing over $m$
for $\langle T_{\tau} \bar{A}_\uparrow^m  (\tau) \rangle_0$
yields
\begin{eqnarray}
\langle S_\sigma \rangle_0
&=& \sum_{m=0}^\infty \frac{1}{m!}
  \langle T_{\tau} \bar{A}_\sigma^m  (\tau) \rangle_0 \nonumber \\
& \equiv &\langle T_\tau
 {\rm exp} \{-\int_0^\beta d \tau \int_0^\beta d \tau'
   \lambda^\sigma (\tau,\tau') {a_\sigma}^\dagger (\tau)
    a_\sigma (\tau') \}  \rangle_0
\nonumber \\
&=& {\rm exp} \{ \sum_n {\rm ln} [1-\lambda_n^\sigma
g_\sigma^0 (i\omega_n)]  \},
\label{seS1}
\end{eqnarray}
with
\begin{equation}
g_\sigma^0 (i\omega_n)=\frac{1}{i\omega_n-(E_\sigma-\mu)}.
\label{seG1}
\end{equation}

The second term provides a time-dependent ``potential''
[the exponential in (\ref{HF})]  which affects the dynamics of
${\bar{A}^n_{-\sigma}(\theta)}$ of the $-\sigma$ electrons.
This term essentially changes the external field
$\lambda_{-\sigma}(\theta, \theta')$ to
$\lambda_{-\sigma}(\theta, \theta') {\rm exp} \{ U
(\theta-\theta') \}$.  Summing over $n$ for
${\bar{A}^n_{-\sigma}(\theta)}$, one needs
to calculate an expectation value  of the type
\begin{equation}
 \langle \tilde{S}_{-\sigma} \rangle_0
\equiv  \langle T_\theta {\rm exp}
         \{-\int_0^\beta d \theta \int_0^\beta d \theta'
   \lambda^{-\sigma} (\theta,\theta')  e^{ U (\theta-\theta')}
   a_{-\sigma}^\dagger (\theta)  a_{-\sigma} (\theta') \} \rangle_0.
\label{seHF}
\end{equation}
These  can not be dealt
with  as in (\ref{seS1})  because of the non-periodicity of
$\lambda^{{-\sigma}}(\theta, \theta') e^{ U
(\theta-\theta')}$.
However,  introducing the  Hamiltonian
$$
\tilde{H}_{-\sigma}=H_0+U a_{-\sigma}^\dagger a_{-\sigma},
$$
and defining  new Heisenberg operators with respect
to $\tilde{H}_{-\sigma}$:
$\tilde{a}_{-\sigma}^\dagger
(\theta)=e^{U\theta}a_{-\sigma}^\dagger (\theta)$,
we  can easily calculate the expectation value
\begin{eqnarray}
\langle \tilde{S}_{-\sigma} \rangle_{\tilde{H}_{-\sigma}}  &=&
\langle  {\rm exp} \{-\int_0^\beta d \theta \int_0^\beta d \theta'
   \lambda^{-\sigma} (\theta,\theta')
   \tilde{a}_{-\sigma}^\dagger (\theta)
   \tilde{a}_{-\sigma} (\theta') \} \rangle_{\tilde{H}_{-\sigma}},\nonumber\\
&=& {\rm exp} \{ \sum_n {\rm ln}
    [1-\lambda_n^{-\sigma} \tilde{g}_{-\sigma}^0 (i\omega_n)] \}.
\label{seHF1}
\end{eqnarray}
Here
\begin{equation}
\tilde{g}_{-\sigma}^0 (i\omega_n)=\frac{1}{i\omega_n-(E_{-\sigma}-\mu)-U},
\label{seG}
\end{equation}
and $\langle \theta \rangle_{\tilde{H}_{-\sigma}}
=Tr e^{\beta (\Omega_{-\sigma}-H_\sigma)} \theta $.
It is clear that
(\ref{seHF1}) describes motion of electrons in the upper Hubbard band,
while (\ref{seS1}) comes from motion in the lower Hubbard band.
This is reminiscent of the alloy analogy approximation (AAA), where
the lower and upper Hubbard bands are introduced explicitly.
However, as we shall see, our approximation avoids the most
serious fault of the AAA and does not violate the Luttinger theroem at
$T=0$, which allows the metallic behavior.

The expectation values of the operators
$\bar{B}_{-\sigma}^n (\theta)$ always lead to a contribution
proportional to $\langle (1-n_\sigma)\rangle$ and a time-dependent
term  $\langle {\rm exp} \{ U\sum_{i=1}^m
(\theta_i-\theta'_i) \} n_{\sigma} \rangle$.  We then
incorporate the exponential
in the latter term directly into an operator $\tilde{A}_{-\sigma}^m$
$= \bar{A}_{-\sigma}^m (\theta) {\rm exp} \{ U\sum_{i=1}^m
(\theta_i-\theta'_i) \}$.
Our decoupling  procedure for
this term is equivalent to writing for  any function
$F[n_{-\sigma}(\theta)]$
\begin{eqnarray}
&&T_\theta \{ e^{\beta (\Omega_0-H_0)} \tilde{A}_{-\sigma}^m (\theta)
F[n_{-\sigma}(\theta)] \} \nonumber \\
&\equiv & T_\theta \{ e^{\beta (\tilde{\Omega}_{-\sigma}-\tilde{H}_{-\sigma}) }
\tilde{A}_{-\sigma}^m (\theta) e^{\beta U n_{-\sigma} }
F[n_{-\sigma}(\theta)]\}
e^{\beta (\Omega_0-\Omega_{-\sigma}) } \nonumber \\
&\approx & \langle T_{\theta} \tilde{A}_{-\sigma}^m (\theta)
      \rangle_{\tilde{H}_{-\sigma}}
 \langle  e^{\beta Un_{-\sigma} } F[n_{-\sigma} (\theta)]
                         \rangle_{\tilde{H}_{-\sigma}}
         e^{\beta (\Omega_0-\Omega_{-\sigma})}.
\label{X1}
\end{eqnarray}
We can see that the upper Hubbard band is introduced
naturally at the decoupling stage.
This procedure must be carried out for $\bar{B}_{\sigma}^n (\tau) $
as well.

 The final result for the decoupling of
the $X_{mn}$ in (\ref{X}) is
\begin{eqnarray}
X_{mn}(\tau, \theta)
    & \approx & \langle T_{\tau}  \bar{A}_\uparrow^m (\tau)\rangle_0
           \langle 1-n_\uparrow \rangle_0
\langle T_{\theta}  \bar{A}_\downarrow^n (\theta)\rangle_0
           \langle 1-n_\downarrow\rangle_0
 \nonumber \\
&& +  \langle T_{\tau}  \tilde{A}_\uparrow^m (\tau)
          \rangle_{\tilde{H}_\uparrow}
           \langle   e^{\beta Un_\uparrow }
   (1-n_\uparrow)\rangle_{\tilde{H}_\uparrow}
         e^{\beta (\Omega_0-\Omega_\uparrow)}
     \langle T_{\theta} \bar{A}_\downarrow^n (\theta)\rangle_0
           \langle  n_\downarrow\rangle_0
 \nonumber \\
&&  + \langle T_{\tau} \bar{A}_\uparrow^m (\tau)\rangle_0
           \langle  n_\uparrow\rangle_0
           \langle T_{\theta} \tilde{A}_\downarrow^n (\theta)
      \rangle_{\tilde{H}_\downarrow}
 \langle  e^{\beta Un_\downarrow } (1-n_\downarrow
                         \rangle_{\tilde{H}_\downarrow}
         e^{\beta (\Omega_0-\Omega_\downarrow)}
\nonumber \\
&& +   \langle T_{\tau} \tilde{A}_\uparrow^m (\tau)
          \rangle_{\tilde{H}_\uparrow}
           \langle   e^{\beta Un_\uparrow }
    n_\uparrow\rangle_{\tilde{H}_\uparrow}
     \langle T_{\theta} \tilde{A}_\downarrow^n (\theta)
      \rangle_{\tilde{H}_\downarrow}
   \langle  e^{\beta Un_\downarrow }  n_\downarrow
                          \rangle_{\tilde{H}_\downarrow}
         e^{\beta \sum_\sigma (\Omega_0-\Omega_\sigma)},
\label{approxX}
\end{eqnarray}
where $\Omega_\sigma$ is the thermodynamic potential of
$\tilde{H}_\sigma$.  Eq.\ (\ref{approxX})
is the main approximation we make for  the
partition functional.

This approximation  leads us to a form for
$\langle S \rangle_0$
\begin{eqnarray}
\langle S \rangle_0 &=&
\langle S_\uparrow\rangle_0 [1-\langle n_\uparrow\rangle_0] \,
\langle S_\downarrow\rangle_0 [1-\langle n_\downarrow\rangle_0]
\nonumber\\
&& + \langle \tilde{S}_\uparrow\rangle_{\tilde{H}_\uparrow}
  [1-\langle n_\uparrow\rangle_{\tilde{H}_\uparrow}] \,
\langle S_\downarrow\rangle_0 \langle n_\downarrow\rangle_0
   e^{\beta (\Omega_0-\Omega_\uparrow)}
\nonumber \\
&&+ \langle S_\uparrow\rangle_0 \langle n_\uparrow\rangle_0 \,
\langle \tilde{S}_\downarrow\rangle_{\tilde{H}_\downarrow}
  [1-\langle n_\downarrow\rangle_{\tilde{H}_\downarrow }]
   e^{\beta (\Omega_0-\Omega_\downarrow)}
\nonumber \\
&&+ \langle \tilde{S}_\uparrow\rangle_{\tilde{H}_\uparrow}
e^{\beta U}
  \langle n_\uparrow\rangle_{\tilde{H}_\uparrow} \,
\langle \tilde{S}_\downarrow\rangle_{\tilde{H}_\downarrow}
e^{\beta  U}
  \langle n_\downarrow\rangle_{\tilde{H}_\downarrow }
   e^{\beta \sum_\sigma (\Omega_0-\Omega_\sigma)},
\label{Z1}
\end{eqnarray}
where $\langle S_\sigma \rangle_0$ and
$\langle \tilde{S}_\sigma \rangle_{H_\sigma}$
 is given by (\ref{seS1}) and (\ref{seHF1}), respectively.
The expectation value of the term
$\langle n_\uparrow n_\downarrow S \rangle_0$ in (\ref{e10}) is
then given by
\begin{equation}
\langle n_\uparrow n_\downarrow S \rangle_0
=\langle \tilde{S}_\uparrow\rangle_{\tilde{H}_\uparrow}
e^{\beta U}  \,
  \langle n_\uparrow\rangle_{\tilde{H}_\uparrow}
\langle \tilde{S}_\downarrow\rangle_{\tilde{H}_\downarrow}
e^{\beta U }
  \langle n_\downarrow\rangle_{\tilde{H}_\downarrow }
   e^{\beta  \sum_\sigma (\Omega_0-\Omega_\sigma)}.
\label{Z2}
\end{equation}

Using (\ref{e10}), (\ref{Z1}) and (\ref{Z2}), we obtain
the partition functional
\begin{equation}
Z=  \left( \begin{array}{ll} \bar{Z}_\uparrow & \tilde{Z}_\uparrow
\end{array} \right)
\left( \begin{array}{ll}  (1-\bar{n}_\uparrow) (1-\bar{n}_\downarrow) &
\bar{n}_\uparrow (1-\tilde{n}_\downarrow)  \\
(1-\tilde{n}_\uparrow)  \bar{n}_\downarrow &
e^{\beta U} \tilde{n}_\uparrow \tilde{n}_\downarrow
\end{array} \right)
\left( \begin{array}{l} \bar{Z}_\downarrow  \\ \tilde{Z}_\downarrow
\end{array} \right) .
\label{e23}
\end{equation}
Here
\begin{eqnarray}
\bar{Z}_\sigma  & =  &
{\rm exp} \{\sum_n {\rm ln} [i\omega_n-(E_\sigma-\mu)-
      \lambda_n^\sigma]  \} ,
\nonumber  \\
\tilde{Z}_\sigma  & =  &
 {\rm exp} \{ \sum_n {\rm ln} [i\omega_n-(E_\sigma-\mu)-U-
              \lambda_n^\sigma] \},
\nonumber\\
\bar{n}_\sigma & \equiv & \langle n_\sigma \rangle_0
={[e^{\displaystyle{\beta(E_\sigma-\mu)}}+1]}^{-1}
\\
\tilde{n}_\sigma & \equiv &\langle n_\sigma \rangle_{\tilde{H}_\sigma}
={[e^{\displaystyle{\beta(E_\sigma-\mu+U)}}+1]}^{-1}.
\nonumber
\label{e24}
\end{eqnarray}
The partition functional $Z$ in  (\ref{e23}) has a very natural
interpretation.  ${\bar{n}_\sigma}$ and ${\bar{Z}_\sigma}$ can
be regarded, respectively, as
the  ``occupancy'' and the ``partition
function'' of the spin $\sigma$ particles in the lower
Hubbard band,
while ${\tilde{n}_\sigma}$ and ${\tilde{Z}_\sigma}$ are the corresponding
quantities in the upper Hubbard band.
The terms   with
${(1-\bar{n}_\uparrow) (1-\bar{n}_\downarrow) }$ ,
${\bar{n}_\uparrow (1-\tilde{n}_\downarrow)}$
(or ${(1-\tilde{n}_\uparrow)  \bar{n}_\downarrow}$ )
and  ${e^{\beta U} \tilde{n}_\uparrow  \tilde{n}_\downarrow}$
in the matrix
account for holes, singly-occupied and
doubly-occupied configurations, respectively.
The factor ${e^{\beta U} }$ is
essential because it cancels the double counting of the interaction $U$
involved in ${\tilde{Z}_\uparrow \tilde{Z}_\downarrow}$.
Comparing the form of the ground state
energy $E$ of the Gutzwiller
approximation in (\ref{EGutzwiller}) with $Z$ in (\ref{e23}), we see that
the ``configuration''  matrix in $Z$ gives a  function like the renormalized
constant $q_\sigma$ in $E$.

For the ${D^{\infty}}$ Falicov-Kimball model, where  one of the
external fields,  ${\lambda^\downarrow}$ say, equals  zero,
Eq.\ (\ref{e23})  reduces to
\begin{equation}
 Z_{FK}=\bar{Z}_\uparrow + \tilde{Z}_\uparrow {\rm exp} [-\beta
  (E_\downarrow -\mu)] ,
\label{FK}
\end{equation}
which is just the result obtained by Brandt and Mielsch \cite{BM}.
Therefore, our partition functional naturally includes the exact
solution of the Falicov-Kimball model as a special case.
It is also easy to check that (\ref{e23}) gives the exact result for the
atomic ($t=0$) and $U=0$ limits.

\section{Thermodynamic properties}

\subsection{The Green's function and physical quantities}

Using   (\ref{e23})  and  (\ref{e6}), one can obtain  the
self-energy functional.  Substituting this functional into the
right hand side of the
self-consistent equation in (\ref{e8}), and noting from
(\ref{se5}) that the left hand side can be written
\begin{equation}
{G^{\rm loc}_\sigma(i \omega_n)}
=[i \omega_n -(E_\sigma-\mu)-
 \lambda_n^\sigma-\Sigma_\sigma (i\omega_n) ]^{-1},
\label{G1}
\end{equation}
one can determine the external fields
self-consistently.  Consequently, the self-energy and
the Green's function for the ${D^{\infty}}$ HM
are  easily found  using this  approximation.

If the external fields are fixed, it is  convenient to use another
form of the local Green's function
derived directly from the functional derivative in (\ref{se5}),
\begin{equation}
{G^{\rm loc}_\sigma(i \omega_n)}=  \bar{\cal G}_\sigma (i\omega_n)
P_0 + \tilde{\cal G}_\sigma (i\omega_n) P_1
\label{G2}
\end{equation}
where
\begin{eqnarray}
\bar{\cal G}_\sigma (i\omega_n)&=&
          {[i\omega_n-(E_\sigma-\mu)-\lambda_n^\sigma]}^{-1},
\nonumber \\
\tilde{\cal G}_\sigma (i\omega_n)&=&
          {[i\omega_n-(E_\sigma-\mu)-U-\lambda_n^\sigma]}^{-1},
\label{G3}
\end{eqnarray}
and
\begin{eqnarray}
P_0 &= &\frac{1}{Z}[\bar{Z}_\uparrow
(1-\bar{n}_\uparrow) (1-\bar{n}_\downarrow) \bar{Z}_\downarrow
  + \bar{Z}_\uparrow
   \bar{n}_\uparrow (1-\tilde{n}_\downarrow) \tilde{Z}_\downarrow],
\nonumber\\
P_1&=& \frac{1}{Z}[\tilde{Z}_\uparrow
(1-\tilde{n}_\uparrow)  \bar{n}_\downarrow  \bar{Z}_\uparrow
  +\tilde{Z}_\uparrow
e^{\beta U} \tilde{n}_\uparrow \tilde{n}_\downarrow \tilde{Z}_\downarrow].
\label{P}
\end{eqnarray}
Clearly, $P_0+P_1=1$. Again,
the Green's function  ${G^{\rm loc}_\sigma(i \omega_n)}$ looks similar
to that given by the alloy analogy approximation \cite{Czy}.
However, $P_0$ and $P_1$ have quite different forms.  We will see
later that this difference leads to important consequences.

Using  (\ref{G2}), the density states of the system can be expressed as
\begin{equation}
{D_\sigma(i \omega_n)}=  \bar{\cal D}_\sigma (i\omega_n)
P_0 + \tilde{\cal D}_\sigma (i\omega_n) P_1
\label{seD}
\end{equation}
where
\begin{eqnarray}
\bar{\cal D}_\sigma (i\omega)&=&-\frac{1}{2\pi i}
[\bar{\cal G}_\sigma (\omega+i0^+)-\bar{\cal G}_\sigma (\omega-i0^+)],
\nonumber \\
\tilde{\cal D}_\sigma (i\omega_n)&=&-\frac{1}{2\pi i}
[\tilde{\cal G}_\sigma (\omega+i0^+)-\tilde{\cal G}_\sigma (\omega-i0^+)].
\label{seD1}
\end{eqnarray}
The chemical potential should be determined self-consistently from the
following equation for the electron concentration
\begin{equation}
n_\sigma(T)= \int_{-\infty}^\infty d \epsilon f(\epsilon)
  D_\sigma(\epsilon),
\label{sen}
\end{equation}
where $f(\epsilon)$ is the fermion distribution function and
$D_\sigma(\epsilon)$ is given in (\ref{seD}).

{}From the Green's function (\ref{G2}), one can calculate
all thermodynamic quantities of the systems.  In this paper
we discuss the suceptibility, specfic heat and resistivity of
the system.

  The  static susceptibility can  be obtained  directly by
differentiating  the occupancy, $n_\sigma$, in
(\ref{sen}) with respect to $E_\sigma$.
We set ${E_\sigma = - \sigma \mu_B h}$, with $h$  the external
magnetic field.
The susceptibilty, $\chi(T)$, is then
\begin{equation}
\chi(T)=\frac{\partial [\mu_B(n_\uparrow-n_\downarrow)]}
  {\partial h} \mid_{h=0}
   =-2\mu_B ^2 (\frac{\partial n_\uparrow}{\partial E_\uparrow}-
  \frac{\partial n_\uparrow}{\partial E_\downarrow} )
  \mid_{E_\uparrow=E_\downarrow=0}
\end{equation}

 In order to obtain the specific heat, it is convenient (especially
for  numerical calculations) to
start from the internal energy, $E(T)$.  This enables us to calculate
the specific heat $C(T)$ as the first derivative of $E(T)$
with respect to $T$ rather than the second derivative of the
free energy.  In the infinite-dimensional case,  $E(T)$
is given by  \cite{PCJ}
\begin{equation}
E(T)=- \sum_\sigma \int d \epsilon \rho_0 (\epsilon)
     \int \frac{d \omega}{2 \pi} f(\omega)
         (\epsilon+\omega) {\rm Im} \frac {1}
      {\omega+\mu-\Sigma_\sigma^r (\omega) -\epsilon + i \delta}
      + \frac{1} {2} \mu n,
\end{equation}
where $\rho_0 (\epsilon)$ is the bare density of states.

The conductivity is expressed  using Kubo's formula as a
current-current  correlation function.
Usually this correlation function can not be calculated exactly
since the vertex corrections are difficult to treat.  However,
in the infinite $D$ limit, vertex corrections
vanish \cite{Khurana}, so that one
is left with a simple bubble for the conductivity correlation function.
The conductivity becomes\cite{PCJ}
\begin{equation}
\sigma(0)=2\int_{-\infty}^\infty  d \epsilon \rho_0(\epsilon)
\int_{-\infty}^\infty \frac{d \omega}{2\pi} (-\frac{\partial f(\omega)}
{\partial \omega}) A(\epsilon, \omega)^2,
\label{cond}
\end{equation}
where  the spectral function $A(\epsilon, \omega)$ is defined as
\begin{eqnarray}
A(\epsilon, \omega)&=&i [G_r(\epsilon, \omega)-G_a(\epsilon, \omega)]
\nonumber\\
&=& i [\frac{1}{\omega+\mu-\Sigma_r+i \eta -\epsilon}
- \frac{1}{\omega+\mu-\Sigma_a-i \eta -\epsilon} ],
\end{eqnarray}
with $\Sigma_r$ and $\Sigma_a$ the retarded and advanced self-energy
of the system, respectively.  We have set $\hbar=e=a=1$ ($a$ is
the lattice constant) in (\ref{cond}).

\subsection{The  Lorentzian bare density of states}

To illustrate the scheme   we use a
Lorentzian type  for the bare DOS \cite{GK, Si}:
\begin{equation}
\rho_0(\epsilon)=\frac{\Gamma}{\pi} (\epsilon^2+\Gamma^2).
\end{equation}
Although this DOS has some undesirable properties as a result of having too
much weight at large energies, it leads to some
analytical results and  some analytical expressions
for physical quantities.
Apart from the Mott transition,  the  physical properties are expected to
be qualitatively correct for this DOS \cite{Si}.

For the Lorentzian DOS, Eq.\ (\ref{e8})  reduces to
\begin{equation}
G^{\rm loc}_\sigma (i\omega_n) = [i\omega-(E_\sigma-\mu) -
\Sigma_\sigma (i \omega_n) +
i\Gamma {\rm sgn} \omega_n]^{-1}.
\label{e26}
\end{equation}
Comparing this with Eq.\ (\ref{G1}),  one immediately obtains
the external field
\begin{equation}
\lambda_n = -i\Gamma {\rm sgn} \omega_n.
\label{e27}
\end{equation}
Substituting  (\ref{e27}) into (\ref{G2}) and (\ref{e6}) gives
the Green's function ${G^{\rm loc}_\sigma(i \omega_n)}$ and
the self-energy ${\Sigma(i \omega_n)}$  for a system with a Lorentzian DOS.
A  frequency summation then gives
$\bar{Z}_\sigma$ and $\tilde{Z}_\sigma$ in (\ref{e23})
\begin{eqnarray}
\bar{Z}_\sigma &=& C \, e^{\beta u_\sigma(E_\sigma-\mu)}, \nonumber \\
\tilde{Z}_\sigma &=& C \, e^{\beta u_\sigma(E_\sigma+U-\mu)},
\label{seZZ}
\end{eqnarray}
where $C$ is a constant, which is irrelevant for the thermodynamic
properties,  and
the function
\begin{equation}
u_\sigma(x)=\int_0^{-x} d \lambda  \int_{-\infty}^\infty f(z-\lambda)
\rho_0(z) dz.
\end{equation}
At  zero temperature, $\bar{Z}_\sigma$  in
(\ref{seZZ})  reduces
\begin{equation}
\bar{Z}_\sigma=C \, {\rm exp}  \big\{
\beta\{\frac{\mu-E_\sigma}{2}+\frac{1}{\pi}
  [(\mu-E_\sigma-U) {\rm arctan}\frac{\mu-E_\sigma}{\Gamma}
-\frac{\Gamma}{2} {\rm ln} (1+\frac{\mu-E_\sigma)^2}{\Gamma^2} ) ] \} \big\},
\label{seZ1}
\end{equation}
while $\tilde{Z}_\sigma$ has the same form as $\bar{Z}_\sigma$ but with
$E_\sigma$ replaced by $E_\sigma+U$.

We find that for the electron concentration $n$ ($=n_\uparrow+n_\downarrow$)
less than one,
the term with ${(1-\bar{n}_\uparrow) (1-\bar{n}_\downarrow) }$ in
$Z$ of (\ref{e23})  is dominant, and $P_1$ vanishes.  The
chemical potential can easily be obtained  from (\ref{sen})
\begin{equation}
\mu=-\Gamma {\rm tan}\frac{\pi}{2}(1-n) ~~~ ({\rm for} ~~ n<1).
\label{nl1}
\end{equation}
For $n>1$, the  term with
${e^{\beta U} \tilde{n}_\uparrow  \tilde{n}_\downarrow}$
in $Z$ is dominant (so that $P_1 =1$) and the chemical potential is given by
\begin{equation}
\mu=U+\Gamma {\rm tan}\frac{\pi}{2}(n-1) ~~~ ({\rm for} ~~ n>1).
\label{nl2}
\end{equation}
For $n=1$, the terms with
${\bar{n}_\uparrow (1-\tilde{n}_\downarrow)}$
[and ${(1-\tilde{n}_\uparrow)  \bar{n}_\downarrow}$ ] in $Z$  are important
and the chemical potential  $\mu=U/2$, as it should be.
The chemical potential $\mu$ is shown in Fig.\ 1 as a function of
the electron concentration $n$.

{}From (\ref{e6}) we find that the  imaginary part of the
self-energy at $\omega=0$ is equal to zero
away from half-filling and the Luttinger theorem is satisfied.
The system is in a Fermi-liquid state.  This is consistent with the
exact statement of Georges and Kotliar \cite{GK}.
However,  at half-filling the imaginary part of  the self-energy at
$\omega=0$ is  not equal to zero and there is a discontinuity
at the chemical potential, as shown in Fig.1.
At half-filling  the
system is therefore
a Mott insulator  for any finite $U$.
This is similar to  what happens in
the one-dimensional case \cite{LW}, but   is
inconsistent with  Monte Carlo calculations \cite{MT}, which
show that the Mott insulating state appears for all $U \geq 3$.
The occurrence of a Mott insulating state at half-filling follows in our
approximation as a result of the particle-hole symmetry and is not
an artefact of using the Lorentzian density of states.
The absence of a Mott transition at  finite $U$
in our approximation results from  overcounting of
the interaction effect for small $U$, {\it i.e.} for the same
reason as given by Hubbard
for his Hubbard-II approximation \cite{Hubbard-II}.

The susceptibilty is given by
\begin{equation}
\chi=2 \mu_B^2 [\frac{P_0}{\pi} \frac{\Gamma}{\mu^2+\Gamma^2}
  + \frac{P_0}{\pi} \frac{\Gamma}{(\mu-U)^2+\Gamma^2} ],
\end{equation}
which has no
divergence for any $n$ and  finite $U$,
so that there is no ferromagnetic phase.

The  temperature dependence of the static
susceptibility ${\chi}$ and the specific heat $C$  are shown in Fig.2.
(The temperature $T$ and the interaction $U$
are in units of $\Gamma$)\,
At high temperature, ${\chi}$ shows a Curie-Weiss-like behavior
(${\chi \sim 1/(T+\theta)}$ with ${\theta >0}$), indicating that the
system behaves like a system of independent moments with
antiferromagnetic fluctuations.
For $n$ near  half-filling or  for large $U$,
${\chi}$ decreases as $T$ drops below a characteristic
temperature, ${T^\star}$, in a way reminiscent of the Kondo effect
(see curves A and B).  Near $T^\star$, the specific heat also  has a peak.
For the susceptibility there appears to be
 a second  temperature ${T_1^\star}$,
below which  a correlated state is formed
where the susceptibility is  nearly  constant, {\it i.e.}, Pauli-like.
At ${T_1^\star}$ there is no anomaly in the specific heat.
For small $n$, the susceptibility continuously crosses over from the
Curie-Weiss behavior to the Pauli-like behavior without
any Kondo-like anomaly.
Correspondingly the specific heat shows no peak.
At very low temperature, the specific heat as a function
of $T$ is consistent with a power law, but with an exponent not
exactly equal to one, {\it i.e.} the behavior is not exactly linear.
The case of the Lorentzian density of states corresponds to
the Anderson impurity model \cite{GK} with $\epsilon_d=\mu(T)$, where
$\epsilon_d$ is the d-level energy.  The impurity model has been
solved exactly for fixed $\epsilon_d$ and does give $C_v \propto T$.
However, in our case the chemical potential $\mu$ is a function of temperature,
which must be determined self-consistently.  This has not yet been done.
The variation of $\mu$ with $T$ will change the impurity model result
for $C_v$.  It is therefore not clear whether our result is actually correct
or a result of the approximation.

  In order to understand the ``Kondo anomaly''  in this system,
we calculate the density of states, shown in Fig.\ 3 for
the case of $U=2$ and $n=0.8$.
There are two peaks above a certain temperature, which corresponds
to the  ``Kondo''  characteristic temperature $T^\star$.
There is only one peak below this temperature.
The Kondo anomaly seems to be related to the appearance of the
Hubbard pseudo-gap.
The self-consistent perturbation calculations also show  one
peak in the DOS of the system  at zero temperature \cite{Muller2, Muller3}.
However, using  the perturbation
scheme of Yosida and Yomada \cite{YY},
Georges and Kotliar \cite{GK} found that the DOS has
two peaks  near half-filling at zero temperature.
The one-peak
feature  found at very low temperature obtained
using the self-consistent pertubation calculation
might indicate that the upper band
does not contribute much to thermodynamic properties at low temperature
and the Hubbard pseudo-gap  behavior manifests itself only above
a characteristic temperature.

The temperature dependence of the resistivity $\rho$ ($=1/\sigma$)
is shown in Fig.\ 4 (for $U=2$) and
Fig.\ 5 (for $U=4$) for various electron concentrations.
The resistivity goes to zero when temperature decreases to zero as
it should, given that our approximation satisfies the Luttinger
theroem at zero temperature.
This is in contrast with the result of the alloy analogy approximation
(shown shematically by the dashed curve in Fig.\ 4), which
keeps rising as temperature decreases. The resistivity becomes
roughly linear with $T$ above the characteristic temperature,
while it rises rapidly around the characteristic temperature.
In order to see clearly how fast $\rho$ changes with $T$, we show in Fig.\ 6
the first derivative of $\rho$ with respective to $T$ in the
case of $U=2$.  There is a  sharp peak at the characteristic temperature.

\section{Discussion and conclusion}

In this paper we have derived an approximate
self-energy functional for the
infinite-dimensional Hubbard  model.
It  naturally includes  the exact solution of
the ${D^\infty}$ Falicov-Kimball model  as a special case
and retains the spin symmetry.   This
finite-temperature theory
successfully incorporates the
high-temperature uncorrelated behavior and the strongly correlated
behavior  at low temperature in a unified way.
Many approximations used before, such as the alloy analogy
approximation,  the equation-of-motion decoupling, the extension of the
Gutzwiller approach and the slave boson mean-field theory, can not
reproduce both limiting behaviors correctly without introducing
a spurious phase transition \cite{Temp} \cite{Czy}.
It seems to us that our approach  is the first to
give physically reasonable results for all
temperatures.

 The functional derived here
is very easy to use.  It can also  be
directly applied to the periodic Anderson model.
The only change is to replace the self-consistent equation
in (\ref{e8}) by
\begin{equation}
G_{f\sigma}^{\rm loc}=
\int_{-\infty}^{\infty}
\frac{\rho_0(\epsilon) d\epsilon}
{i\omega_n-(\epsilon- \mu)-\Sigma_{f \sigma} (i\omega_n)
- \displaystyle{\frac{V^2}{i\omega_n - (\epsilon-\mu)} } },
\label{loc8}
\end{equation}
where  ``$f$'' represents localized electrons, and
$V$ is the mixing energy between conduction
and localized electrons.  From (\ref{loc8}) one can
determine the  local Green's function and the
self-energy of the localized electrons, and
consequently, all Green's functions of the
system \cite{GKS}.

In this paper we  have treated  only the case of a
homogeneous system.   For some symmetry-broken
phases, such as the antiferromagnetic state, the
self-consistent equation  changes  \cite{GKS},
although the form of self-energy functional
remains.  Becuase our  scheme treats spin-up and spin-down
electrons on an equal footing (in contrast to the alloy
analogy approximation), we expect that it can describe correctly
the antiferromagnetic instability.  We will
discuss this issue elsewhere \cite{Li2}.

There exist some weaknesses in the present form
of the partition functional.  The dynamic flucuations
induced by the motion of one spin species to another
are not properly treated.  Moreover, the Mott transition
does not take place at a finite interaction $U$.
Since our approximation is an expansion,
we can include relevant higher order contributions which account for
these flucuations.  The Mott
transition has been produced by Hubbard in his Hubbard-III
paper \cite{Hubbard-III} by including
the resonant correction in the equivalent coherent-potential
approximation.  It would be interesting to examine this transition
in our formalism by including  higher order fluctuation effects
since our solution satisfies the
Luttinger theorem in the metalic side at zero temperature, while the
Hubbard-III solution does not \cite{Edwards}.

\medskip
\begin{center}
ACKNOWLEDGMENTS
\end{center}

YML  would like to thank  Z. B. Su, T. Xiang and  L. Yu and
for helpful discussions.
We acknowledge  support from  the SERC of the United Kingdom
under grant No.\ GR/E/79798, and from MURST/British Council under grant
No.\ Rom/889/92/47.

\newpage

\begin{center}
FIGURE CAPTIONS
\end{center}

\begin{description}
\item{Fig.1.} The chemical potential $\mu$ as a function
  of the electron concentration $n$ at zero temperature. $\mu$ is
  in unit of $\Gamma$.  The gap at half-filling is present for all
       non-zero $U$
\item{Fig.2.} The temperature dependence of the susceptibility, ${\chi}$
      (solid curve), and the specific heat $C$ (dashed curve) for the
       Lorentzian density of states
     for various $n$ and $U$. Curves
      $A$ $\&$ $a$ are for  $U=1$ and $n=0.9$,
      while $B$  $\&$ $b$, $C$ $\&$ $c$ and $D$ $\&$ $d $ correspond to
      $n=0.8$, $0.6$ and
       $0.4$, respectively, for the case of $U=2$.
      $T$ and $U$ are in units of $\Gamma$, and $\chi$ is in unit of
      $\mu_B^2$.
\item{Fig.3.}  The density of states of the system, $D(\omega)$, for
   various temperatures in the case of
       $U=2$ and $n=0.8$.
\item{Fig.4.}  The temperature dependance of the resistivity
     for $U=2$ and various electron concentration $n$.
    The result of the alloy analogy approximation is
  schematically shown by the dashed curve.
\item{Fig.5.}  The temperature dependance of the resistivity
       for $U=4$ and  various $n$.
\item{Fig.6.} The first derivative of the resistivity with respect to
   temperature  for  $U=2$, corresponding to Fig.\ 4.
\end{description}

\end{document}